\documentclass[aps,prl,twocolumn,noshowpacs]{revtex4-1}
\usepackage{amsmath}
\usepackage{color}
\usepackage{graphicx}
\usepackage{epstopdf}

\usepackage [english]{babel}
\usepackage [latin1]{inputenc}

\begin{document}

\title{Ordered structure of FeGe$_2$ formed during solid-phase epitaxy}

\author{B.~Jenichen}
\email{bernd.jenichen@pdi-berlin.de}
\author{M.~Hanke}
\author{S.~Gaucher}
\author{A.~Trampert}
\author{J.~Herfort}
\affiliation{Paul-Drude-Institut f\"ur Festk\"orperelektronik, Leibniz-Institut im Forschungsverbund Berlin e.V.,
Hausvogteiplatz 5--7, D-10117 Berlin, Germany}

\author{H.~Kirmse}
\author{B.~Haas}
\affiliation{Humboldt-Universit\"at zu Berlin, Institut f\"ur Physik,
Newtonstra{\ss}e 15, D-12489 Berlin,  Germany}

\author{E.~Willinger}
\author{X.~Huang}
\affiliation{Fritz-Haber-Institut der Max-Planck-Gesellschaft, Faradayweg 4, D-14195 Berlin, Germany}

\author{S.~C.~Erwin}
\affiliation{Center for Computational Materials Science,
Naval Research Laboratory,
Washington, DC 20375, USA}

\date{\today}

\begin{abstract}
Fe$_{3}$Si/Ge(Fe,Si)/Fe$_{3}$Si thin film stacks were grown by a combination of molecular beam epitaxy and  solid phase epitaxy (Ge on Fe$_{3}$Si). The stacks were analyzed using electron microscopy, electron diffraction, and synchrotron X-ray diffraction. The Ge(Fe,Si) films crystallize in the well oriented, layered tetragonal structure FeGe$_{2}$ with space group P4mm . This kind of structure does not exist as a bulk material and is stabilized by solid phase epitaxy of Ge on Fe$_{3}$Si. We interpret this as an ordering phenomenon induced by minimization of the elastic energy of the epitaxial film.
\end{abstract}



\maketitle

Ordering phenomena of epitaxial layers have been found in semiconductor mixed
crystals as well as in metallic alloys. In general, the ordering has a strong
influence on the physical properties of the epitaxial films.
In semiconductors (SC), the formation of monolayer superlattices in mixed crystal
Al$_x$Ga$_{1-x}$As epitaxial films grown by metal-organic chemical vapor deposition
on (110) or (100) oriented  GaAs substrates has been observed.\cite{Kuan1985}
The authors suggested that this long-range ordering is a thermodynamically stable
phase at temperatures below about 800~$^\circ$C. A strain-induced order-disorder
transition was found in SiGe epitaxial films grown on Si(001) by molecular beam
epitaxy (MBE).\cite{Ourmazd1985} This phenomenon later was explained using
self-consistent total energy calculations.\cite{Martins1986,Zunger1987}

For metallic alloys, the amount of collected data is even larger.\cite{Gorsky1928,Willliams1935,Khachaturyan1973,Khachaturyan1983,Ruban2008,Zhuravlev2014,Wrobel2015}
Here, the influence of ordering on material properties like hardness, conductivity,
magnetism and corrosion resistance is important. The elastic interaction of the
different atoms of the alloys often leads to energetically favored ordered structures.
This kind of ordering is influenced by the anisotropy of the crystal lattice.

The structures of the epitaxial Ge and Fe$_{3}$Si films on GaAs substrates
correspond well to the known structures of their bulk materials.\cite{tinkham08,Jenichen09,jenichen2005}
However, when the Fe$_{3}$Si film is used as a substrate for epitaxial growth of Ge,
the influence of the Fe$_{3}$Si structure on the growing epitaxial Ge film unexpectedly
turns out to be stronger and ordering phenomena occur. These ordering phenomena are
induced by the epitaxial growth and were not observed in bulk material up to now.
Several methods were applied to achieve perfect semiconducting Ge films on top of
ferromagnetic (FM) layers.\cite{Yamada2012,Jenichen2014a,Kawano2016}  Recently,
the method of solid-phase epitaxy (SPE) of Ge was utilized in order to achieve a
perfect crystallinity of the film and superior interface quality.\cite{Gaucher2017,Sakai2017,Kawano2017}
However, the diffusion of Fe and Si was not entirely prevented during the annealing process.
Therefore, the Ge film contained some amount of Fe and Si, leading to a shift of
the X-ray diffraction (XRD) peak of the Ge(Fe,Si) film and the formation of a
superlattice-like structure inside the Ge(Fe,Si) film. The FM Fe$_{3}$Si forms
Schottky contacts with the SC Ge and GaAs.\cite{Hamaya2013} A triple layer
structure FM-SC-FM is therefore suitable for Schottky barrier tunneling transistors
described in \cite{Wu2005}, similar to tunneling magneto-resistance devices.\cite{Yuasa2007,oogane2011}
A spin dependent transport of holes was detected up to room temperature.\cite{Kawano2017}
The aim of the present paper is the investigation of the structure of the Ge(Fe,Si) film.

Fe$_{3}$Si/Ge(Fe,Si)/Fe$_{3}$Si thin film stacks were grown combining MBE for Fe$_{3}$Si
on GaAs(001) and SPE for Ge on Fe$_{3}$Si.\cite{Gaucher2017} A 36~nm thick Fe$_{3}$Si
film was grown by MBE on the GaAs buffer layer at a growth rate of 16~nm/h and a
temperature of 200~$^\circ$C in a separate growth chamber dedicated to metal growth.
In the same chamber the 4~nm thick Ge film was deposited at 150~$^\circ$C resulting
in a smooth interface but with an amorphous structure. For the SPE of the Ge film
the sample was heated at 5~K/min up to a temperature of 240~$^\circ$C and then
annealed for 10~min. The 12~nm thick upper Fe$_{3}$Si film was then grown by MBE on
top of the crystalline Ge under the same conditions as the lower Fe$_{3}$Si film.
The growth and annealing conditions of the sample result in a typical structure
characteristic for the whole series.\cite{Gaucher2017} After sample preparation
transmission electron microscopy (TEM) and  XRD (here at an energy of E~=~10keV)
were used for structural characterization. Experimental details are given in the
supplemental material.\cite{supplement2018} TEM and XRD simulations were performed
using software packages available.\cite{stadelmann2016,CrystalMaker2017,Stepanov1997}
In addition, density functional theory (DFT) was employed for the calculation of the
lattice parameter and the electronic band structure of the Ge(Fe,Si). DFT in the
generalized gradient approximation~\cite{Perdew1996} was applied using the
Vienna Ab Initio Simulation Package.\cite{Kresse1996A,Kresse1996B} The
Perdew-Burke-Ernzerhof (PBE)~\cite{Perdew1996a} and the Heyd-Scuseria-Ernzerhof (HSE)~\cite{Heyd2003}
exchange-correlation functionals, were used for the calculations.

From earlier X-ray results it is clear that the diffusion inside the layer stack
has an obvious influence on the formation of the structure of the Ge(Fe,Si) film.\cite{Gaucher2017}
Here the diffusion during SPE is more important than the diffusion during the subsequent epitaxial
growth of Fe$_{3}$Si, because the characteristic structure was observed even without
the uppermost Fe$_{3}$Si film, and the diffusion during Fe$_{3}$Si film growth is known to be low.\cite{Herfort2006}
We obtained the depth dependence of the atomic composition of the
different elements by energy dispersive X-ray (EDX) spectroscopy in the scanning
TEM (STEM).\cite{supplement2018} The Ge(Fe,Si) film consisted of a Ge content
of 60$\pm$5 at\%, an Fe content of 35$\pm$5 at\% and a Si content of 5$\pm$5 at\%.\cite{foot2018}
Considering in a first approximation the binary phase diagram of Fe--Ge, the phases
of FeGe, and FeGe$_{2}$ could be expected for the given composition range and an
annealing temperature of 240~$^\circ$C during the SPE process.\cite{Jaafar2010} According
to such a consideration the FeGe$_{2}$ should have the tetragonal Al$_{2}$Cu
structure (I~4/m~c~m).\cite{Satya1965}

 Let us consider the formation of our Ge(Fe,Si) thin film structure in more detail.
 During SPE, an initially amorphous material is annealed on top of a crystalline
 substrate resulting in a lattice-matched crystalline epitaxial film. In a solid
 solution inside the growing film, at first sight a random distribution of the
 elements on the different lattice sites can be expected. However, an ordered
 distribution of the solute atoms can sometimes lead to a minimum of the free
 energy F of the system. The distributions of the different elements can be
 described in the static concentration-wave formalism.\cite{Khachaturyan1973,Khachaturyan1983}
 A heterogeneity $\Delta({\overrightarrow{r}})$ can be written as

\begin{equation}
\Delta({\overrightarrow{r}})~=~[n({\overrightarrow{r}})-n_0]
\end{equation}

where $n({\overrightarrow{r}})$ is the occupation probability of a lattice site
with a certain type of atom, $n_0$ is the average concentration of that element,
and ${\overrightarrow{r}}$ is the site-vector of the lattice in the crystalline
film. The concentration-wave representation of the heterogeneity $\Delta({\overrightarrow{r}})$ is
written as follows:
If all the positions of the crystal lattice sites are described by one Bravais
lattice $\Delta({\overrightarrow{r}})$ can be expanded in a Fourier series, i.e.
it can be considered as a superposition of static concentration waves:

\begin{equation}
\Delta(\overrightarrow{r}) = \frac{1}{2}\sum_j^[Q(\overrightarrow{k_j})exp(i{\overrightarrow{k_j}~\overrightarrow{r}})+Q^*(\overrightarrow{k_j})exp(-i{\overrightarrow{k_j}~\overrightarrow{r}})]
\end{equation}

where $Q(\overrightarrow{k_j})$ is the static concentration wave amplitude and
can be treated like a long-range order parameter, and $\overrightarrow{k_j}$ is
the nonzero wave-vector of the static concentration wave defined in the first
Brillouin zone of the disordered alloy. The index $j$ denotes the wave vectors
in the Brillouin zone. The ordering can result in a reduction $\Delta$F of the free
energy. Then the uniform solid solution becomes unstable with respect to the
heterogeneity~(2) with a certain concentration wave
vector ${\overrightarrow{k}}={\overrightarrow{k}_0}$. In our epitaxial films we
clearly observe such an ordering.

In our experiment, the interface between the underlying Fe$_{3}$Si film and an
amorphous Ge layer is the starting point of the SPE. The lattice mismatch between
Ge and Fe$_{3}$Si is $\Delta{a}/a~=~1.5\cdot10^{-4}$. During the deposition of the
Ge and the subsequent annealing, Fe and Si atoms diffuse into the Ge film, leading
to a small but finite lattice mismatch. This lattice mismatch can be compensated
not only by a tetragonal distortion of a disordered Ge(Fe,Si) film, but in addition
by an ordering of a substitutional solid solution which can be described as a
concentration wave with the wave vector $\overrightarrow{k}_0$. From symmetry
considerations, it is clear that the wave vector of the static concentration
wave $\overrightarrow{k}_0$ should be perpendicular to the Fe$_{3}$Si/Ge(Fe,Si)
interface, leaving the properties of the film unchanged along the interface.
And indeed, the experimental results obtained earlier by XRD and TEM showed
the formation of a superlattice only along one direction, the direction perpendicular
to the Fe$_{3}$Si/Ge(Fe,Si) interface.\cite{Gaucher2017}    A possible choice for
the length of the vector is $|\overrightarrow{k}_0|~=~2\pi/c$, where $c$ is the
superstructure period observed. At the same time we take $c$ as the lattice parameter
of the growing Ge(Fe,Si) lattice perpendicular to the interface, and $a$ as the
lattice parameter parallel to the interface. Such a choice of the lattice leads
us to the possibility, that the ordering can be described in the frame of the
Ge(Fe,Si) lattice itself with a basis of two types of lattice sites described by
fractional lattice coordinates: one type occupied mainly by Ge-atoms (or Si-atoms)
and the other mainly by Fe-atoms. In this case we can
write $(\overrightarrow{k}_0 \cdot \overrightarrow{r}) = 2\pi z$
where $z$ is the coordinate  perpendicular to the Fe$_{3}$Si/Ge(Fe,Si) interface.
The occupation probability $n(z)$ for a certain type of atom is then
\begin{equation}
n(z) = (1/2)\cdot \eta \cdot cos (2\pi z/c) + n_0
\end{equation}
where $\eta$ is the order parameter and $n_0$ is the average concentration.

\begin{figure}[!t]
\includegraphics[width=8.5cm]{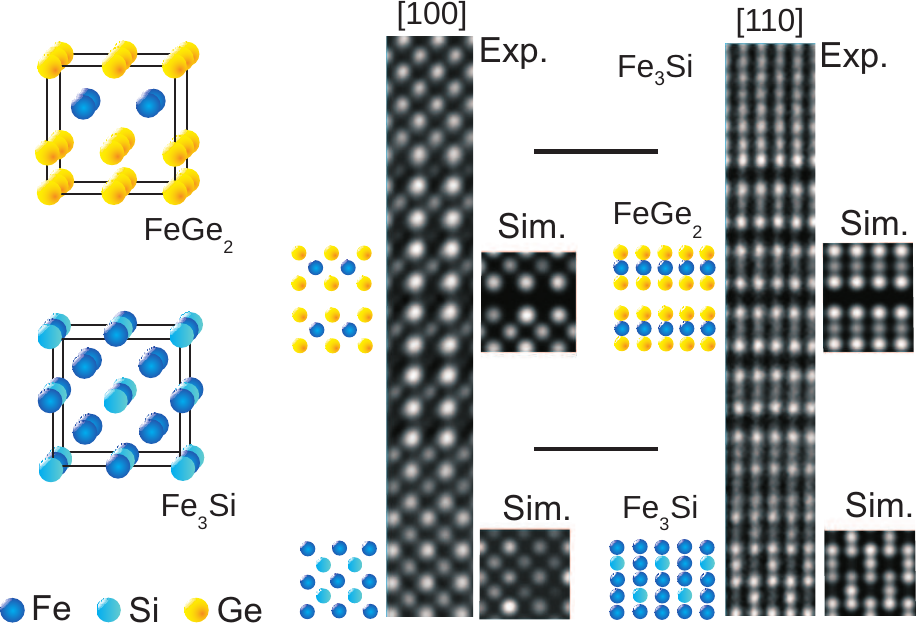}
\caption{(color online) Comparison of the HAADF experimental cross-section
micrographs (larger rectangles, Exp.) with the structural models of FeGe$_{2}$ P4mm
and Fe$_{3}$Si shown on the left side as well as the corresponding simulations
(small squares, Sim.). The structure of Fe$_{3}$Si is well known, whereas the
structure of FeGe$_{2}$ is obtained from the Z-contrast of the present micrographs,
taken along the two projections [100] and [110]. The horizontal lines mark the
positions of the FeGe$_{2}$/Fe$_{3}$Si interfaces. They are 1~nm long.
}
\label{fig:comparison}
\end{figure}

\begin{figure*}[!t]
\includegraphics[width=10.0cm]{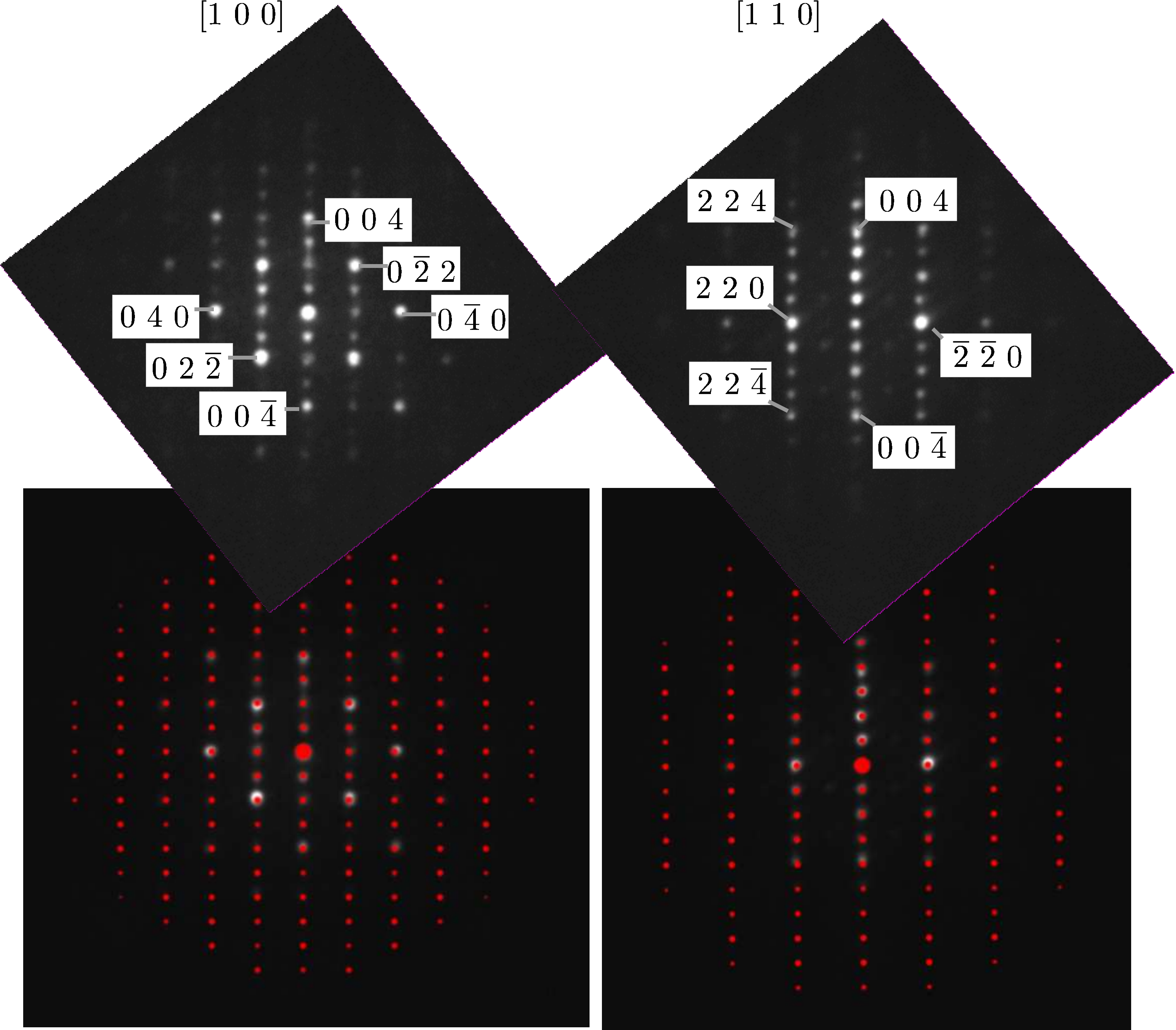}
\caption{(Color online) Nano-beam diffraction patterns of the thin FeGe$_2$ film from [100] (left) and [110] (right) oriented samples. The comparison of the experimental patterns with the results of the simulations (red) is given below.
}
\label{fig:nbd}
\end{figure*}

A calculation of the change of the free energy $\Delta$F would need more detailed
information about the structure of the film. That is why for further investigation
of the structure we performed Z-contrast imaging in the STEM. The Z-contrast mode
is an incoherent imaging method. In a first approximation presuming constant thickness
and neglecting the influence of strain, a high angle annular dark field (HAADF) STEM
micrograph exhibits Z-contrast: The intensity diffracted by an atomic column is I$_{HAADF}~\sim~Z^{1.7...2}$,
thus heavier atoms give brighter image contrast.\cite{Pennycook1990} The intensity
increases with the number of atoms in a column as well.\cite{VanAert2011} The STEM
micrographs where evaluated using the method of template-matching using the symmetry
in the growth plane.\cite{Zuo2014} Original data is presented in the supplemental material.\cite{supplement2018}

In Fig.~\ref{fig:comparison}, we can recognize the superstructure in the Ge-rich Ge(Fe,Si) film.
The image of the [100] oriented sample shows brighter spots forming a square lattice.
These spots are caused by Ge-columns (Z=32). The darker spots, which occur as center of
every second square are due to Fe-columns (Z=26). In the image of the [110] oriented
sample we see rows of brighter spots and can attribute them to Ge-columns. Between
every second pair of bright rows we recognize darker spots and consider them as contrasts
due to Fe-columns.
In the Fe$_{3}$Si film, we recognize the typical Fe-triplets of the image of the [110]
oriented sample and the faint spots of the Si-columns between them (Z=14). The image of
the [100] oriented sample shows a square lattice of relatively bright spots with darker
spots in the centers of the squares. The \textit{DO$_3$} structure of Fe$_{3}$Si
corresponding to this kind  of contrast is known and can serve as a reference.
On the basis of the Z-contrast of our HAADF micrographs obtained along the two crystal
orientations~[100]~and~[110], we are able to propose a structural model for the Ge(Fe,Si)
film: It is the FeGe$_{2}$ (P4mm) structure shown on the left side of Fig.~\ref{fig:comparison}.
Four unit cells are depicted for better correspondence with the Fe$_{3}$Si lattice. The
structural models of Fe$_{3}$Si (below, given as a reference) and FeGe$_{2}$ (above, our
proposal) are drawn. The experimental micrographs are compared to the structural models
giving an illustration of our proposal of the FeGe$_{2}$ structure. On the other hand,
the well-known structure of the Fe$_{3}$Si films is well reproduced, and so we can be sure
that we described the FeGe$_{2}$ structure in a good approximation.

The verification of the proposed FeGe$_{2}$ structure (see Fig.~\ref{fig:comparison}) can
be done using computer simulation of HAADF micrographs. We performed the simulations in
the frozen phonon approximation using the parameters of the probe-C$_s$-corrected
HR-STEM (JEOL ARM200) operating at 200~kV. Simulations and experimental micrographs
shown in Fig.~\ref{fig:comparison} agree well, indicating that a proper structural model
was found. The location of the 5~at\% Si detected by EDX spectroscopy is still unclear.
The Si atoms probably are located on Ge sites. Besides, we found differently ordered regions
of the Fe$_{3}$Si, the \textit{B}2 order located near the interface and the \textit{DO$_3$}
order in depth of the Fe$_{3}$Si film.\cite{hashimoto2007jvst}

\begin{figure}[!t]
\includegraphics[width=10.0cm]{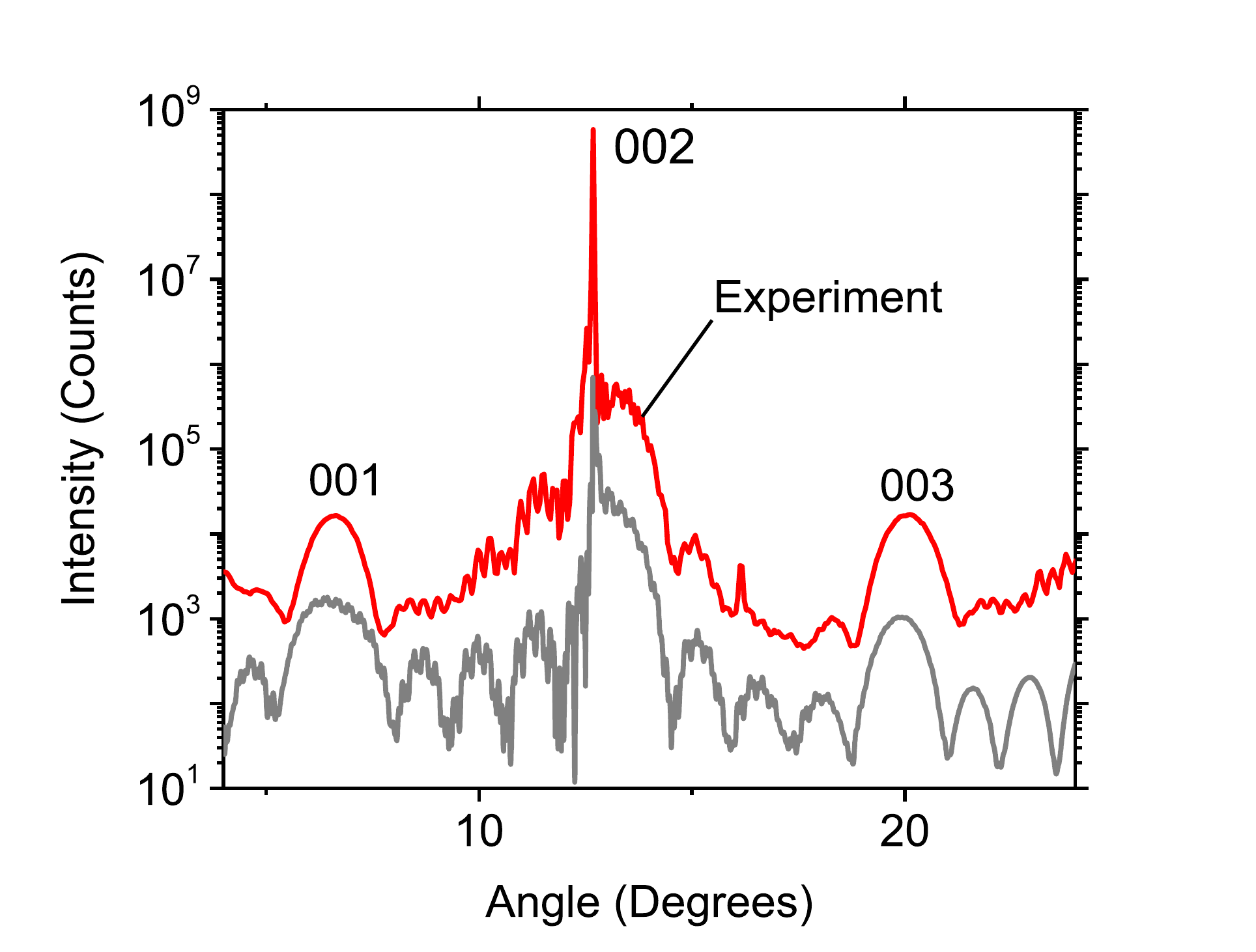}
\caption{(color online)  Comparison of the measured XRD curve and the corresponding simulation
(below) near GaAs(002) for the Fe$_{3}$Si/FeGe$_{2}$/Fe$_{3}$Si film stack on GaAs(001),
obtained using the structure shown in Figure~\ref{fig:comparison}.
}
\label{fig:xrd1}
\end{figure}

In Fig.~\ref{fig:nbd} nano-beam diffraction patterns of the thin FeGe$_2$ film from [100] and [110]
oriented samples are given. The patterns were fully indexed using the proposed FeGe$_2$ structure model
and simulated in kinematical approximation. The results of the simulations given below in red color
agree well with the experiments further supporting our structural model.

\begin{table}
\caption{Experimental lattice parameters (determined by XRD) a of Fe$_{3}$Si,  and 2a
and c of FeGe$_{2}$ films grown on a GaAs(001) substrate in comparison with unstrained
lattice parameters of Fe$_{3}$Si and FeGe$_{2}$ calculated by density functional theory
for PBE and  HSE functionals.} \vspace{12pt} \label{tab:data2}
\begin{tabular}{llll}

~~~~&~EXP~&~PBE~&~HSE\\
\hline

a(Fe$_{3}$Si)~~~ &~0.5654~nm & ~0.561~nm~&~0.575~nm~ \\
2a(FeGe$_{2}$)~~~ &~0.5654~nm &  ~0.572~nm ~&~0.580~nm~\\
c(FeGe$_{2}$)~~~& ~0.5517~nm &~0.544~nm~&~0.549~nm~ \\

\hline

\end{tabular}
\label{tab:tab1}
\end{table}

In Fig.~\ref{fig:xrd1}, the XRD curve (symmetrical $\omega/2\Theta$-scan, i.e. the 00L
crystal truncation rod) together with the simulation
of the diffraction curve of the Fe$_{3}$Si/FeGe$_{2}$/Fe$_{3}$Si film stack in the vicinity
of the GaAs(002) peak are shown. Here, the simulated diffraction curve agrees with the main
features of the experimental diffraction curve, especially the FeGe$_{2}$ 001 and 003 maxima
are visible, and the FeGe$_{2}$ 002 peak is shifted with respect to GaAs 002. In the
supplemental material the XRD reciprocal space map of the non-symmetric 20L crystal
truncation rod is shown.\cite{supplement2018} All relevant diffraction maxima of the
reciprocal space map are positioned on a vertical line perpendicular to the sample
surface, i.e. the structures are elastically strained and no plastic relaxation
occurs.\cite{Heinke1994} From XRD we deduce the lateral lattice parameter of the epitaxial
layer stack a~=~0.5654~nm~=~2~$\cdot$~0.2827~nm and the strained vertical lattice parameter
of the FeGe$_{2}$ thin film of c~=~0.5517~nm (cf. Table~\ref{tab:tab1}). A more careful
analysis of the STEM HAADF micrographs allowed for determination of the strained lattice
plane distances of two types of sublayers in the FeGe$_{2}$ structure, viz. empty and filled
ones. Empty layers and filled layers have distances of c$_1$~=~0.266~nm and and c$_2$~=~0.282~nm
respectively. Filling with Fe leads to an expansion of the distance of the corresponding layer.
This fact points to the possibility of strain compensation between the two sublayers of the FeGe$_{2}$ lattice
as a driving force for the formation of the ordered superlattice-like structure in the
epitaxial layer stack.\cite{jenichen2017footnote3} On the other hand, the integral layer
thickness c$_1$+c$_2$  determined by STEM corresponds well to the strained c-value determined
by XRD for the FeGe$_{2}$ tetragonal lattice as a whole (see above). The theoretical values of
the lattice parameters in Table~\ref{tab:tab1} are given for unstrained lattices. In an
epitaxial layer stack additional tetragonal deformation occurs.\cite{Hornstra1978,foot2018}

\begin{figure}[!t]
\includegraphics[width=10cm]{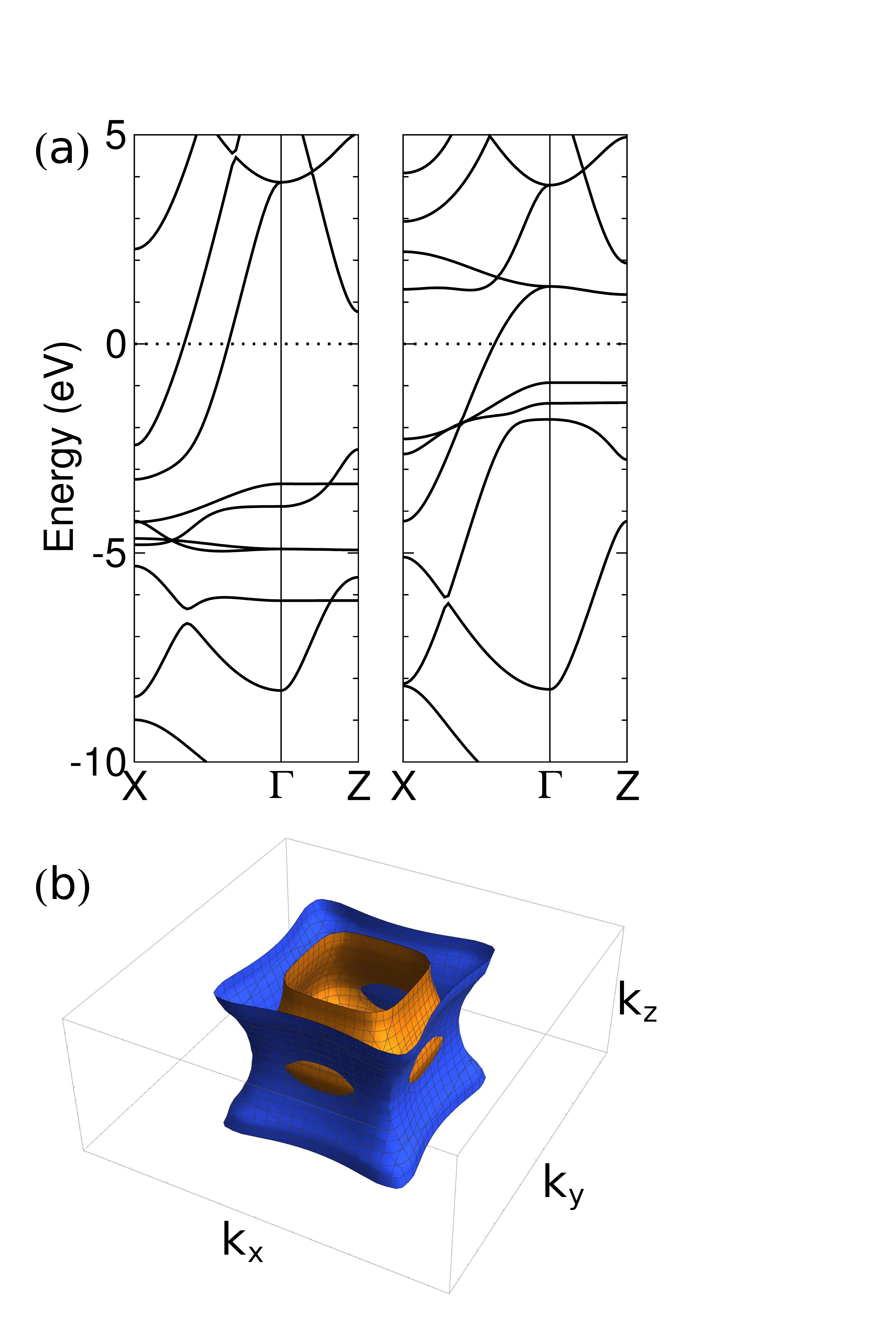}
\caption{(color online) (a) Band structure of FeGe$_2$ in the P4mm structure calculated using
DFT with the HSE screened hybrid functional. Left panel is majority spin; right panel is minority
spin. (b) Fermi surfaces for majority (dark, blue) and minority (bright, gold) spins. A second
sheet in the majority spin channel is not shown.
}
\label{fig:bands_DFT}
\end{figure}

Our results show that the ordering can be considered as a systematic arrangement of Fe-atoms
and -vacancies in a CsCl-type FeGe lattice, where both atoms and vacancies are found on the
Fe-sites, and the number of Fe-atoms is reduced by half in order to obtain the stoichiometry
of FeGe$_{2}$. A random positioning of the Fe-atoms would lead to a cubic lattice. But in our
case we have the boundary condition at the FeGe$_{2}$/Fe$_{3}$Si interface, where the in-plane
lattice parameter of FeGe$_2$ is fixed to a value of 0.2827~nm. Let us take two CsCl-type unit
cells to describe the FeGe$_{2}$ lattice as a result of ordering of the Fe atoms and vacancies.
Then the diffraction intensity of the fundamental 002 reflection is proportional
to $\mid$\textit{f}$_{Fe}$~+~2$\cdot$\textit{f}$_{Ge}$$\mid^2$, where \textit{f}$_{Fe}$
and \textit{f}$_{Ge}$ are the atomic form factors of the Fe-atom and the Ge-atom, respectively.
The intensity of the 001 superlattice reflection is
\begin{equation}
I({\overrightarrow{k}}) ~{\sim}~\mid\eta\cdot~f_{Fe}~-~(1-\eta)\cdot~f_{Fe}\mid^2
\end{equation}

because all other contributions vanish and only the ordered Fe-atoms give a diffraction signal.
From the comparison of the intensities of the layer reflections 001 (superlattice) and 002
(fundamental) we obtain $\eta$ = (0.805$\pm$0.02), i.e the ordering is nearly complete. The
film consists of an almost ideal FeGe$_{2}$ lattice.  From the principle of minimum free energy
the energy \textit{V} lost by an atom moving from a disorder position to an order position can
be calculated in the Gorsky-Bragg-Williams approximation.\cite{Gorsky1928, Willliams1935} This
means in our case
\begin{equation}
V =kT~\cdot~\log[(1~+~\eta)/(1~-~\eta)].
\end{equation}
For an order parameter $\eta$ = 0.805 and our annealing temperature of \textit{T} = 513~K we obtain \textit{V}=~(0.042$\pm$0.02)~eV per atom.

Thanks to the ordered structure of the FeGe$_{2}$ film with the extended Fe-sheets we are
expecting outstanding properties of the new material. As a first step, using the structural
data of the FeGe$_{2}$ obtained in the present work, we have calculated by DFT the band
structure shown in Fig.~\ref{fig:bands_DFT}(a). We can see, that the Fermi surfaces
in Fig.~\ref{fig:bands_DFT}(b) consist of cylinders along the z-axis, i.e. perpendicular
to the Fe-sheets. The electrical properties in the plane of the Fe-sheets probably will
differ considerably from those perpendicular to the sheets. FeGe$_{2}$ belongs to a class
of quasi-two-dimensional materials similar to MoS$_{2}$.\cite{Rotjana2018} Two-dimensional FeGe$_{2}$
can now be fabricated with a thickness down to one nanometer, and high--T$_C$
superconductivity seems to be possible in such a structure.\cite{Stewart2011,Miiller2015,Ge2015,Zhou2018}
Thanks to the well ordered Fe-sheets, the
concentration-waves can be accompanied by spin-density-waves.

Single crystal Ge-rich films were successfully grown by solid phase epitaxy on Fe$_{3}$Si(001).
Surprisingly the structure of the films was not the expected diamond structure of Ge, but a well
oriented and layered tetragonal FeGe$_{2}$ P4mm structure. A lattice misfit caused by
interdiffusion of Si, Fe, and Ge leads to the formation of a new structure and ordering inside
the film. We observe here one of the rare cases, where epitaxy is causing the formation of a
distinct crystal structure differing from the equilibrium bulk structure, in particular the
strain-induced ordering of the FeGe$_{2}$ film with a periodicity along the direction
perpendicular to the FeGe$_{2}$/Fe$_{3}$Si interface.

The authors thank Claudia Herrmann for her support during the
MBE growth, Doreen Steffen, Margarita Matzek and Sabine Krau{\ss} for sample preparation,
and Uwe Jahn for critical reading of the manuscript and helpful discussion.
This work was supported in part by the Office of Naval Research through the
Naval Research Laboratorys Basic Research Program. Some computations were performed
at the DoD Major Shared Resource Center at AFRL. We thank the Helmholtz-Zentrum Berlin (HZB)
for providing beamtime at the BESSY-beamline U125/2  KMC
with the endstation PHARAO.

%

\end{document}